\def\dl{\bf} \catcode`\@=11

\def\a{\alpha}

\def\ad{{\rm ad}\,}

\def\alg           {algebra}

\def\be            {\begin{equation}}
\def\bearl         {\begin{array}{l}}
\def\bearll        {\begin{array}{ll}}
\def\bearlll       {\begin{array}{lll}}

\def\bfe           {{\bf1}}

\def\brev          {\breve}

\def\caij{\hat{a}_{ij}}

\def\ce            {\hat{\epsilon}}
\def\cft           {conformal field theory}
\def\cfts          {conformal field theories}

\def\chii          {\raisebox{.15em}{$\chi$}}

\def\Chi           {{\cal V}}
\def\Chil          {\Chi_\Lambda^{(\omega)}}
\def\chil          {\chii_\Lambda^{[\omega]}}

\def\cI            {\breve{I}}

\def\csa           {Cartan subalgebra}

\def\d{\delta}
\def\D{\Delta}
\def\D+{\Delta^{\!+}_{\phantom{X}}}

\def\diag          {{\rm diag}}

\def\dyd           {Dynkin diagram}
\def\e{\epsilon}
\def\ee            {\end{equation}}
\def\eE            {{\rm e}}

\def\eear          {\end{array}}

\def\epsh          {\hat\epsilon}

\newcommand\fcft[3]{{{#1}^{\mskip-#3 mu\raise #2 pt\hbox{$\scriptstyle\circ$}}}}
\def\findim        {finite-dimensional}
\newcommand\fline[1]{\vfill\noindent ------------------\\[1 mm]}

\def\futnot#1      {\ifnum\draftcontrol=1
                   \footnote{~{\sc internal footnote:} #1}\ \fi}
\def\futnote#1     {\footnote{~#1}\ }
\def\g             {{\sf g}}

\def\gb            {\mbox{$\bar\g$}}

\def\gO            {\mbox{$\brev{\mbox{\g}}$}}

\def\hil           {\mbox{$\cal H$}}

\def\hl            {\mbox{${\cal H}_\Lambda$}}

\def\hlo           {\mbox{${\cal H}_{\om^\star\Lambda}$}}

\def\hy            {$\mbox{-\hspace{-.66 mm}-}$}

\def\ii            {{\rm i}}

\def\iN            {\!\in\!}

\def\J             {J}

\def\kma           {Kac\hy Moo\-dy algebra}
\def\Kma           {Kac--Moo\-dy algebra}

\long\def\labl#1   {\label{#1}\ee}

\def\lie           {Lie algebra}
\def\Lie           {Lie group}
\def\LL            {\Lambda}

\def\modinv        {modular invarian}

\def\o             {\omega}
\def\oaii          {\sum_{l=0}^{N_i-1}\!a_{i,\omD^li}}
\def\oaij          {\sum_{l=0}^{N_j-1}\!a_{i,\omD^lj}}

\def\ocha          {twining character}

\def\olie          {orbit Lie algebra}
\def\om            {\omega}
\let\omchar=\ocha
\newcommand\omd[1] {{\dot\omega #1}}
\def\omD           {\dot\omega}

\newcommand\omt[1] {{\omega^\star #1}}

\def\omtLa         {{\omt\Lambda}}

\def\oW            {\hat W}

\long\def\query#1{\hskip 0pt{\vadjust{\everypar={}\small\vtop to 0pt{\hbox{}%
     \vskip -13pt\rlap{\hbox to 50.0pc{\hfil{\vtop{\hsize=8pc\tolerance=6000%
     \hfuzz=.5pc\rightskip=0pt plus 3em\noindent#1}}}}\vss}}}}%

\def\rep           {representation}
\def\Rep           {Representation}

\newcommand\sect[1] {\section{#1}\setcounter{equation}{0}}
\newcommand\Sect[2]{\sect{#1}\label{s.#2} \ifnum\draftcontrol=1 \query{s.#2}\fi}

\def\tauo          {{\tau_\omega}}

\def\trhl          {{\rm tr}^{}_{{\cal H}_{\Lambda}}}

\def\U             {{\sf U}}

\newcommand\version[1] {\ifnum\draftcontrol=1 \typeout{}\typeout{#1}\typeout{}
                   \vskip3mm \centerline{\fbox{\tt DRAFT -- #1 -- \today}}  
                   \vskip3mm \fi}

\def\wh            {\hat w}
\def\What          {\hat W}

\def\WO            {\brev{W}}

\def\wzwts         {WZW theories}

\def\zet           {{\dl Z}}

\input{psbox.tex} \let\fillinggrid=\relax

\catcode`\@=12

\documentstyle[epsf,12pt]{article}

\begin{document}

\begin{flushright}  {~} \\[-23 mm] {\sf hep-th/9612060} \\
{\sf UCB-PTH-96/56}, \,  {\sf LBNL-39653} \\
{\sf NIKHEF 96-030}, \,  {\sf IHES/M/96/79}
\\[1 mm]{\sf November 1996} \end{flushright} \vskip 2mm

\begin{center} \vskip 12mm

{\Large\bf TWINING CHARACTERS AND} \vskip 0.3cm
{\Large\bf ORBIT LIE ALGEBRAS $^\dagger$}
\vskip 12mm
{\large J\"urgen Fuchs} $^1$\, , {\large Urmie Ray} $^2$\, ,
{\large Bert Schellekens} $^3$
\,\ and \,\ {\large Christoph Schweigert} $^4$
\\[9mm] {$^1$ \small DESY, Notkestra\ss e 85,\, D -- 22603~~Hamburg}
\\[2mm] {$^2$ \small IHES, 35, route de Chartres, F -- 91440~~Bures-sur-Yvette}
\\[2mm] {$^3$ \small NIKHEF/FOM, Kruislaan 409, NL -- 1098 SJ~~Amsterdam}
\\[2mm] {$^4$ \small Department of Physics, University of California, 
         Berkeley, CA 94720, USA, and}
\\ {\small Theoretical Physics Group, Lawrence Berkeley National 
         Laboratory,}
\\{\small Berkeley, CA 94720, USA}
\end{center}

\vskip 15mm

\begin{quote} {\bf Abstract.} \\
We associate to outer automorphisms of generalized \Kma s generalized 
character-valued indices, the {\em twining characters}. A character
formula for twining characters is derived which shows that they coincide 
with the ordinary characters of some other generalized \Kma, the so-called
{\em orbit Lie algebra}. Some applications to problems in \cft, 
algebraic geometry and the theory of sporadic simple groups are sketched. \\
\end{quote}

\vfill {}\fline{} {\small
$^\dagger$~~ Slightly extended version of a talk given by C. Schweigert at the 
   XXI International Colloquium on Group Theoretical Methods in Physics
   (Goslar, Germany, July 1996) \newpage

\section{Generalized \Kma s}

Generalized \kma s constitute a class of \lie s which comprise many \lie s that
describe symmetries in physical systems. In particular, they include
the \findim\ simple \lie s (i.e.\ the four series of classical \lie s
and the five exceptional simple \lie s), twisted and untwisted affine \lie s
(i.e.\ centrally extended loop algebras), hyperbolic \lie s (such as $E_{10}$),
as well as the Monster \lie\ and its various relatives. Moreover, any chiral
algebra of a \cft, i.e.\ any vertex operator algebra, gives
rise to a generalized \kma. 

Any \findim\ simple \lie\ is generated (as a \lie) by several 
copies of $\rm sl(2)$, one copy for each simple root. This is also true for
ordinary \kma s; a generalized \kma, however, is generated by copies of 
$\rm sl(2)$ and of the Heisenberg algebra 
which has a basis $\{e,f,K\}$ with $K$ a central element and non-trivial Lie 
bracket $[e, f] = K$.
Nonetheless, generalized \kma s can still be characterized 
by a square matrix, the Cartan matrix
$A=(a_{ij})_{i,j\in I}$. The index set $I$ can either be
finite, $I=\{1,2,...\,,n\}$, or countably infinite,
$I=\zet_+$. In case of \findim\
simple \lie s the entries of $A$ are integers; here we allow for real 
entries, but still we keep the following properties of $A$:
\be\begin{array}{ll}
{\rm (i)} \, & a_{ij}\leq 0\;$ if $\,i\not=j; \qquad 
{\rm (ii)} \ \ {{2a_{ij}}\over {a_{ii}}} \in \mbox{\zet}
\ \ \mbox{if}\ \ a_{ii}>0; \\[1.2mm]
{\rm (iii)} & \mbox{if}\ \ a_{ij}=0, \; \mbox{then}\ \ a_{ji}=0; \\[1.2mm]
{\rm (iv)} &  \mbox{there exists a diagonal matrix 
              $D=\diag(\e_1,... ,\e_n),$  with
              $\e_i$ a positive} \\[.2mm]
           & \mbox{real number for all $i$, such that $DA$ is symmetric.}
\end{array}\ee

{}From these data the generalized \kma\ is constructed by the same procedure 
that is used to construct a \findim\ simple \lie\ from its Cartan matrix. One 
starts with an abelian \lie\ $\mbox{\g}_0$ of dimension greater or equal 
to $n$ and fixes $n$ linearly independent elements $h_1,...\,,h_n$ of 
$\mbox{\g}_0$ and $n$ linear forms $\a_j$ on $\mbox{\g}_0$, the simple roots, 
such that $\a_j(h_i)=a_{ij}$.
The generalized \kma\ $\mbox{\g}=\mbox{\g}(A)$ with Cartan matrix $A$ and 
Cartan subalgebra $\g_0$ is then the \lie\ generated by $e_i,\, f_i\,$, 
$i\iN I$, and $\mbox{\g}_0$, modulo the relations 
\be \begin{array}{l}
[e_i, f_j]=\d_{ij}h_i \,, \quad [h, e_i]=\a_{i}(h)e_i \,, \quad
[h, f_i]=-\a_{i}(h)f_i \,,  \\[1.9mm]
(\ad e_i)^{1-2{a_{ij}/a_{ii}}}e_j=0=(\ad f_i)^{1-2{a_{ij} / a_{ii}}}
  f_j \quad \mbox{if} \quad  a_{ii}>0 \,, \\[1.9mm]
\mbox{}[e_i, e_j]=0=[f_i, f_j]\quad \mbox{if} \quad a_{ij}=0 \,.
\end{array} \ee

\section{Automorphisms of generalized \Kma s}

We now turn to the main object of our interest: a class of outer
automorphisms of generalized \kma s and some new structures associated to them. 
Such automorphisms occur naturally in many physical applications; for some
examples see the last section and \cite{jcon}.
We start with a permutation $\omD$ of finite order of the index set $I$ which 
leaves the Cartan matrix $A$ invariant:
\be a_{\omD i, \omD j}=a_{i,j} \, . \ee
If the generalized \kma\ has a \dyd, $\omD$ corresponds to a symmetry of the
\dyd. Such a permutation $\omD$ induces an automorphism 
$\omega\!:\, \mbox{\g}\to\mbox{\g}$ of the \lie\ \g,
which is defined by its action on the generators $e_i$, $f_i$ and $h_i$:
\futnote{In general, one also must describe the action of $\omega$ on the 
derivations of the generalized \Kma. For more details see section 3 of 
\cite{fusS3}.}
\be   \omega(e_i) :=  e_{\omd i}\, ,  \quad \omega(f_i) :=  f_{\omd i} \, ,  
\quad \omega(h_i) := h_{\omd i} \, .  \ee

The automorphism $\om$
gives rise to a map $\tauo\!:\; \hl\to\hlo$ of integrable highest weight 
modules (and analogously also for Verma modules) which `$\omega$-twines' the 
action of \g:
  \be  \tauo(R_\Lambda(x)\cdot v)= R_\omtLa(\omega(x))\cdot\tauo(v) \,,  \ee
and maps a fixed highest weight vector $v_\Lambda$ of $\hl$ to a highest 
weight vector $v_{\omega^\star\Lambda}$
of $\hlo$. Analogous maps can also be defined
for Verma modules; they are compatible with the submodule structure of the
Verma modules and hence also with their null vector structures.
Pictorially, the map $\tauo$ can be represented as follows:\\[1em]{}
  $$   \vbox to 210bp{
  \psannotate{\psboxscaled{500}{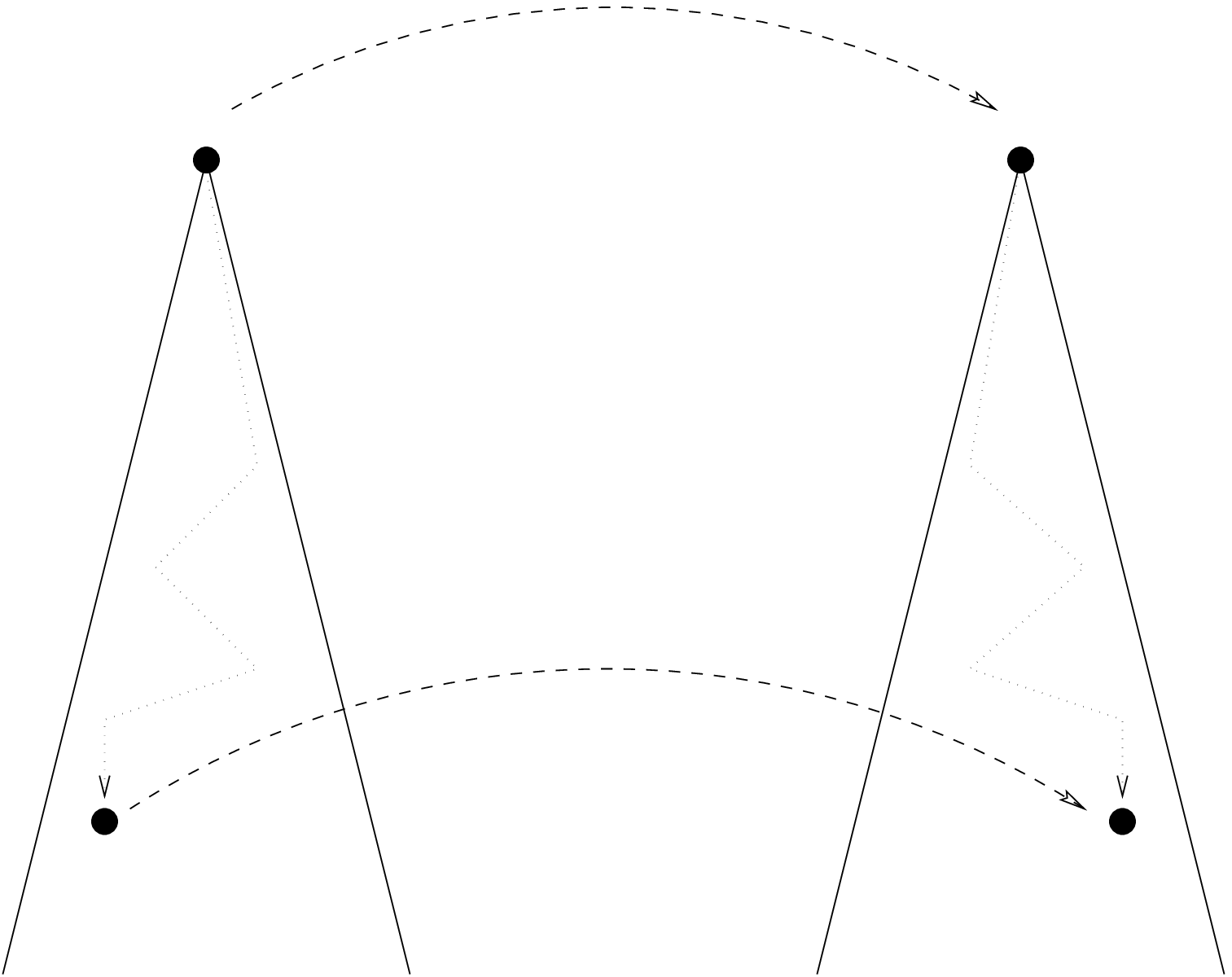}}{\fillinggrid
       \at(4.5\pscm;-1.0\pscm){$\hil_\Lambda$}
       \at(15.0\pscm;-1.0\pscm){$\hil_{\omega^\star\Lambda}$}
       \at(3.4\pscm;1.0\pscm){{\normalsize$v=xv_\Lambda$}}
       \at(12.3\pscm;1.0\pscm)
        {{\footnotesize $\tauo v=\omega(x)v_{\omega^\star\!\Lambda}$}}
       \at(9.0\pscm;4.2\pscm){$\tauo$}
       \at(5.0\pscm;7.5\pscm){{\footnotesize$x\!\in\!\U(\g_-)$}}
       \at(9.1\pscm;7.5\pscm){{\footnotesize$\omega(x)\!\in\!\U(\g_-)$}}
       \at(3.7\pscm;10.0\pscm){$v_\Lambda$}
       \at(13.8\pscm;10.0\pscm){$v_{\omega^\star\Lambda}$}
       \at(7.5\pscm;12.5\pscm){$\tauo$}
  }} \,  $$ 

\section{Fixed points}

Note that $\tauo$ is generically a linear map from a 
module $\hl$ to a different module $\hlo$. 
A particularly interesting situation occurs when the highest weight 
$\Lambda$ is {\em symmetric}, i.e.\ satisfies $\omega^\star(\Lambda)=\Lambda$. 
In this case the heighest weight module $\hl$ is called a {\em fixed point} 
of $\omega$ and $\tauo$ is an endomorphism of the vector space underlying the
module $\hl$. To keep track of some properties of the endomorphism $\tauo$, 
we introduce a character-like object, the {\em \ocha\/} $\chil$:
  \be  \chil(h)= \trhl \tauo\, \eE^{2\pi\ii R_\Lambda(h)} \;\ \mbox{for}\ 
  h\iN \mbox{\g}_0 \, . \ee
Like ordinary characters the \omchar s are (formal) functions on the \csa.
They are the generating functions of the trace of $\tauo$ on the weight spaces,
i.e.\ they are generalized (since $\omega$ does not necessarily have order two)
character-valued indices. Clearly, for the trivial automorphism 
$\omega=\bfe$ we recover the ordinary character of the module $\hl$.

It is a rather surprising result \cite{fusS3,furs} that the \omchar\ is 
essentially 
identical to the character of some other generalized \kma\ 
\mbox{\gO}, called the {\em \olie}. This result makes the twining characters
explicitly computable. In particular, it implies 
that the coefficients in the expansion of the \omchar\
are not arbitrary complex numbers, but non-negative integers,
and hence justifies a posteriori the name twining {\em character}.

The \olie\ corresponding to \g\ and its diagram automorphism $\omega$ is 
obtained by a simple prescription which corresponds to folding the \dyd\ 
of \g\ according to the action of $\omD$. Pictorially we have e.g.:
  $$   \vbox to 160bp{
  \psannotate{\psboxscaled{500}{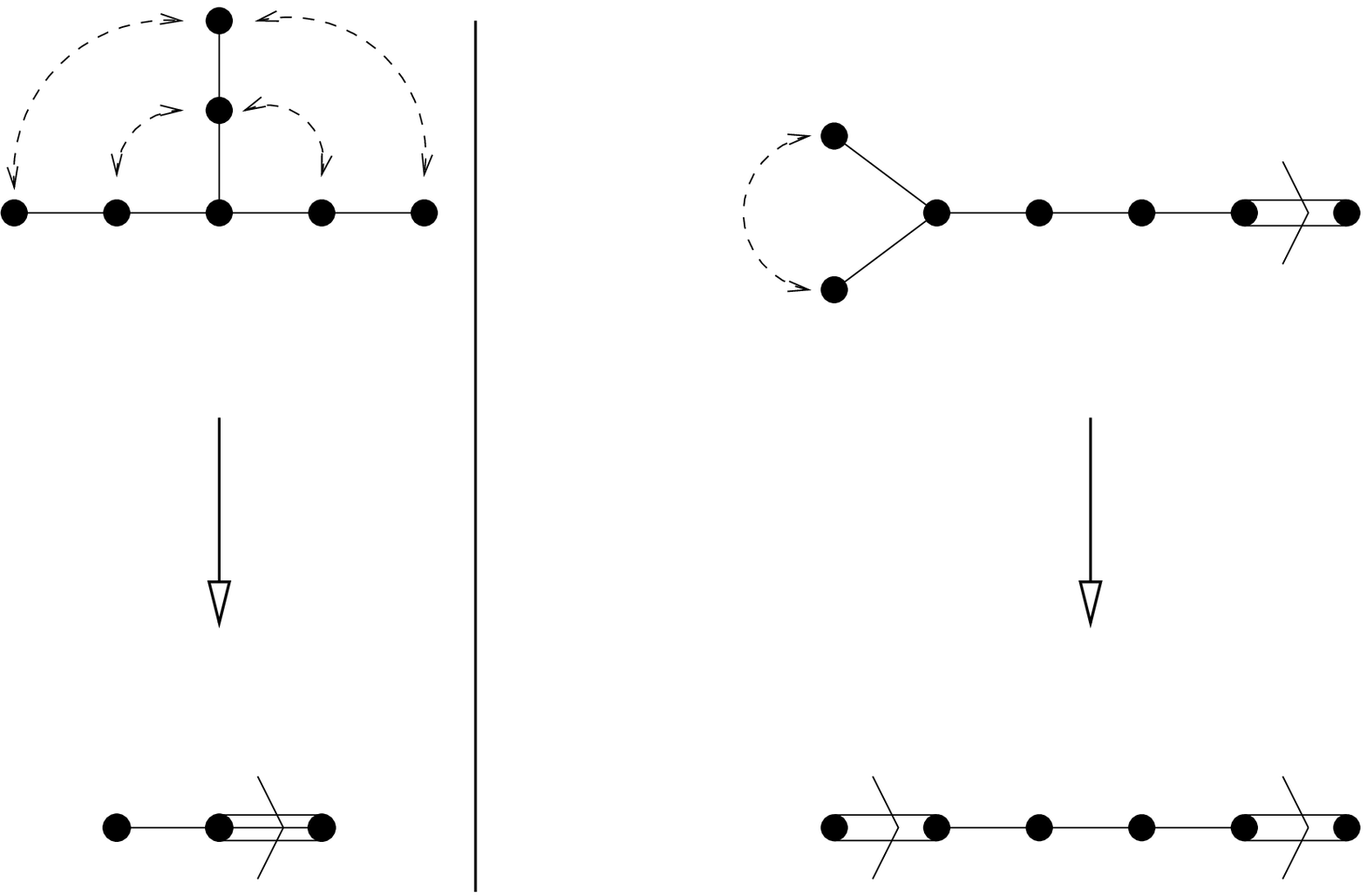}}{\fillinggrid
       \at(-1.0\pscm;0.5\pscm){{\footnotesize $\gO= G_2^{(1)}$}}
       \at(7.5\pscm;0.5\pscm){{\footnotesize $\gO= \tilde B_5^{(2)}$}}
       \at(-3.0\pscm;8.0\pscm){{\footnotesize $\mbox{\g}= E_6^{(1)}$}}
       \at(6.5\pscm;8.0\pscm){{\footnotesize $\mbox{\g}= B_6^{(1)}$}}
  }} $$ 

In formul\ae, the orbit \lie\ is described as follows. Denote by $\hat I$ a set 
of representatives in $I$ for each $\omD$-orbit. The Cartan matrix of the 
\olie\ is then labelled by the subset 
  \be \cI:=\{i\iN{\hat I}\mid\oaii\leq 0\Longrightarrow \oaii=a_{ii}\}  \ee
of $\hat I$. For each $i\iN {\hat I}$ we define the number
 \be s_i:=\cases{{a_{ii}/\oaii}&if $i\iN\cI$ and $a_{ii}\not=0\,,$\cr
  \noalign{\vskip 2pt} \ 1&otherwise$\,$,\cr}  \ee
which is either $1$ or $2$. The Cartan matrix 
${\hat A}=({\caij})_{{i,j}\in {\cI}}$ of the orbit \lie\ is then defined 
by summing over one index of the Cartan matrix of $\g$:
  \be \caij:=s_j\oaij. \ee 
We emphasize that \mbox{\gO} is not constructed as a subalgebra of \g; in 
particular the \olie\ is in general not isomorphic to the subalgebra of \g\ 
that stays fixed under $\omega$. 
One can also show that the algebra \g\ and the 
\olie\ \mbox{\gO} are 
\futnote{except for the order-$N$ automorphisms of ${\rm sl}(N)$}
of the same type (i.e.\ simple, affine, 
indefinite); the \olie\ of hyperbolic \lie s is hyperbolic as well. 
A particularly interesting situation is illustrated
on the right hand side of the figure: for certain \olie s, one obtains the 
twisted affine \lie s $\tilde B^{(2)}_n$ which are the only twisted affine
\lie s for which the characters have good modular transformation properties.
This had been predicted 
\cite{fuSc2} on the basis of level-rank dualities of $N=2$ superconformal 
coset theories. 

We conjecture that when the original \lie\ \g\ is an untwisted affine \lie\
and the symmetry of the \dyd\ corresponds to the action of a simple current,\,%
\futnote{\dyd\ symmetries coming from simple currents \cite{scya6} are 
elements of the unique maximal abelian normal subgroup of the
diagram automorphisms. This subgroup
is isomorphic to the center of the universal covering Lie group
that has the horizontal subalgebra $\gb\subset\g$ as its \lie.}
then the modular matrix $S$ of the \olie\ \mbox{\gO} describes the
modular transformation properties of the 
chiral blocks for the one-point function of the relevant simple current
on the torus.
This would provide a conceptual explanation of the observation that in these 
cases the twining characters of \g\ span again a unitary \rep\ of 
$SL(2,\zet)$. For some further evidence for this conjecture, see \cite{jcon}.

\section{Sketch of the proof}

We will now outline the main ideas which enter in the proof of a character
formula for the \ocha; for the full details we refer the reader to the
original publications \cite{fusS3,furs}.
The first step is to show that the subgroup $\What$ of the Weyl group $W$ of 
\g\ that consists of those elements of $W$
whose action on the weight space of \g\ commutes with $\omega^\star$ is 
isomorphic to the Weyl group $\WO$ of the \olie\ $\gO$. This part of the proof 
also provides new insight in the structure of Coxeter groups.

The idea is to show that $\What$ plays for the \ocha s the role the 
full Weyl group plays for the ordinary characters: 
First, the twining characters of Verma modules are 
$\What$-odd. More precisely, let $\Chil$ denote the twining 
character of the Verma module; then 
$ \Chi^{(\omega)} := \eE^{-\rho-\Lambda} \, \Chi_\Lambda^{(\omega)} $
(which does not depend on $\Lambda$) is antisymmetric under $\What$,
$ \wh(\Chi^{(\omega)}) = \epsh(\wh) \, \Chi^{(\omega)}$.
Here $\epsh$ is the sign function associated to $\What$ as a Coxeter group, 
and not the restriction of the sign function of $W$. 
The twining characters $\chil$ of the irreducible 
modules, on the other hand, are symmetric under the action of $\What$:
\be \wh(\chil) = \chil \, . \ee
We can now generalize the arguments used in the proof of the Weyl-Kac-Borcherds
character formula of ordinary characters and implement the symmetry properties 
under 
$\What$ to derive an explicit character formula for the twining characters:
  \be  \chil = {\sum_{\wh\in\oW}\ce(\wh)\,\wh(S_{\LL}^{\o})\over
  \sum_{\wh\in\oW}\ce(\wh)\,\wh(S_{0}^{\o})}\, , \labl{exfor}
where 
  \be  S_{\LL}^{\omega}=\eE^{\Lambda+\rho}\sum\ce(\beta)\,\eE^{-\beta} \ee
and $\,\ce(\beta)=(-1)^n$ if $\beta$ is the symmetric sum of $n$ pairwise
orthogonal imaginary simple roots, all orthogonal to $\Lambda$, and
$\ce(\beta)=0$ otherwise. 
Comparison with the character formula for the \olie\ then proves the claim.
\\[-1.5em]

\section{Applications and conclusions}

Orbit \lie s and twining characters are novel structures in the \rep\ theory 
of generalized \kma s. They already found several applications
in two-dimensional \cft, e.g.\ in the solution of the problem of field 
identification fixed points in coset \cfts\ \cite{fusS4} and in
fixed point resolution for integer spin simple current (`$D$-type') modular 
invariants of \cfts\ \cite{fusS6}.
More details can be found in the original publications and in \cite{jcon}.

In the case of \wzwts, integer spin modular invariants describe the WZW theory
on a non-simply connected group manifold. In general, twining characters 
and \olie s play a role as soon as one considers non-simply connected \Lie s.
In particular, they allow to describe Chern-Simons theories based on such 
groups \cite{schW3} and they give a Verlinde formula also for non-simply 
connected groups \cite{fusS6}. Other applications concern the biggest sporadic 
simple group, the monster group, which acts on a generalized \kma, the 
monster \lie, by outer automorphisms (recently, this group has been proposed 
as the symmetry underlying $N=2$ string theory). We expect that \olie s
and \ocha s will be a useful tool in this situation as well.
Finally, twining characters of \findim\ simple \lie s are 
closely related to the characters of non-connected \findim\ simple 
\Lie s (the latter have e.g.\ been studied in \cite{shin2}).

\bigskip\bigskip
\small\noindent{\bf Acknowledgement.} 
J.\ Fuchs was supported by a Heisenberg fellowship. C.\ Schweigert
was supported in part by the Director, Office of Energy Research,
Office of Basic Energy Sciences, of the U.S.\ Department of Energy under
Contract DE-AC03-76F00098 and in part by the National Science Foundation
under grant PHY95-14797.

  \newcommand{\wb}{\,\linebreak[0]} \def\wB {$\,$\wb}
  \newcommand{\Bi}[1]    {\bibitem{#1}}
  \newcommand{\Erra}[3]  {\,[{\em ibid.}\ {#1} ({#2}) {#3}, {\em Erratum}]}
  \newcommand{\BOOK}[4]  {{\em #1\/} ({#2}, {#3} {#4})}
  \newcommand{\vypf}[5]  {\ {\sl #5}, {#1} [FS{#2}] ({#3}) {#4}}
  \renewcommand{\J}[5]     {\ {\sl #5}, {#1} {#2} ({#3}) {#4} }
  \newcommand{\Prep}[2]  {{\sl #2}, preprint {#1}}
  \newcommand{\Cont}[2]  {{\sl #2}, {#1}}
 \def\ijmp  {Int.\wb J.\wb Mod.\wb Phys.\ A}
 \def\jmsj  {J.\wb Math.\wb Soc.\wB Japan}
 \def\npbF  {Nucl.\wb Phys.\ B\vypf}
 \def\npbp  {Nucl.\wb Phys.\ B (Proc.\wb Suppl.)}
 \def\nuci  {Nuovo\wB Cim.}
 \def\nupb  {Nucl.\wb Phys.\ B}
 \def\phlb  {Phys.\wb Lett.\ B}
 \def\comp  {Com\-mun.\wb Math.\wb Phys.}
 \def\anop  {Ann.\wb Phys.}
 \def\A       {Algebra}
 \def\alg     {algebra}
 \def\Be     {{Berlin}}
 \def\BIR    {{Birk\-h\"au\-ser}}
 \def\Ca     {{Cambridge}}
 \def\CUP    {{Cambridge University Press}}
 \def\furu    {fusion rule}
 \def\GB     {{Gordon and Breach}}
 \newcommand{\inBO}[7]  {in:\ {\em #1}, {#2}\ ({#3}, {#4} {#5}),  p.\ {#6}}
 \def\Infdim  {Infinite-dimensional}
 \def\NY     {{New York}}
 \def\Q       {Quantum\ }
 \def\qg      {quantum group}
 \def\Rep     {Representation}
 \def\SV     {{Sprin\-ger Verlag}}
 \def\syms    {sym\-me\-tries}
 \def\wzw     {WZW\ }

 \end{document}